\documentclass[a4,12pt]{article}
\usepackage{type1cm,amsmath,amssymb,color,graphics,amscd,amsfonts,mathrsfs,epsf,indentfirst,cite}
\usepackage{epsfig}
\usepackage{bm}
\usepackage{subfigure}
\usepackage{multirow,graphicx}
\usepackage[dvipdfm,hypertex]{hyperref}
\makeatletter
\@addtoreset{equation}{section}

\makeatletter
\renewcommand\section{\@startsection {section}{1}{\z@}%
                                   {-3.5ex \@plus -1ex \@minus -.2ex}
                                   {2.3ex \@plus.2ex}%
                                   {\normalfont\large\bfseries}}
\renewcommand\subsection{\@startsection{subsection}{2}{\z@}%
                                     {-3.25ex\@plus -1ex \@minus -.2ex}%
                                     {1.5ex \@plus .2ex}%
                                    {\normalfont\bfseries}}

\begin{document}


\begin{titlepage}
  \thispagestyle{empty}
  
  \begin{flushright} 
    DAMTP-2011-22 \\
    KUNS-2327
  \end{flushright} 
  
  \vspace{2cm}
  
  \begin{center}
    \font\titlerm=cmr10 scaled\magstep3
    \font\titlei=cmmi10 scaled\magstep4
    \font\titleis=cmmi7 scaled\magstep4
     \centerline{\titlerm
Anisotropic Inflation 
 with Non-Abelian Gauge Kinetic Function}
    
    \vspace{2.2cm}
    \noindent{
     Keiju Murata\footnote[1]{K.Murata@damtp.cam.ac.uk},
     Jiro Soda\footnote[2]{jiro@tap.scphys.kyoto-u.ac.jp}
      }\\
    \vspace{0.8cm}
    
   {\it ${}^1$ DAMTP, University of Cambridge, Centre for Mathematical Sciences,\\
   Wilberforce Road, Cambridge CB3 0WA, UK}\\
   {\it ${}^2$ Department of Physics,  Kyoto University, Kyoto, 606-8502, Japan} 
   \vspace{1cm}
   
   {\large \today}
  \end{center}

  \vskip 3em

\begin{abstract}
We study an anisotropic inflation model with a gauge kinetic function
for a non-abelian gauge field.
We find that, in contrast to abelian models, the anisotropy can be 
either a prolate or an oblate type, which could lead to a different prediction 
from abelian models for the statistical anisotropy
in the power spectrum of cosmological fluctuations.
During a reheating phase, we find chaotic behaviour of the non-abelian gauge field
which is caused by the nonlinear self-coupling of the gauge field.
We compute a Lyapunov exponent of the chaos which turns out to be uncorrelated
with the anisotropy.
\end{abstract}

\end{titlepage}

\section{Introduction}

In an inflationary scenario, quantum vacuum fluctuations during inflation
accounts for the origin of the large scale structure of the universe.
The nature of such primordial fluctuations is understood from the approximate symmetry 
in quasi-deSitter inflation. First of all, in order to have the inflation, 
we need an approximate translation invariance for an inflaton field, 
which forbids strong nonlinearity in the action.
  Hence, we have Gaussian statistics of fluctuations. 
Moreover, the approximate deSitter symmetry makes the power spectrum of fluctuations 
scale invariant and statistically isotropic. Note that
the scale invariance originates from the temporal part of the deSitter symmetry
and the statistical isotropy comes from the spatial part of that. 
Since these predictions are based on the symmetry,
they are robust in the standard single field inflationary scenario.

However, the symmetry in quasi-deSitter inflation is not accurate 
from the point of view of precision cosmology.
In fact, if the translation symmetry is exact, inflation never ends.
Since the violation of the translational symmetry is characterized by the slow roll
parameters, it is natural that the non-Gaussianity
in a single inflaton model is of the order of the slow roll
 parameters~\cite{Maldacena:2002vr}.  
 Apparently, the violation of deSitter symmetry is also characterized
by slow roll parameters of the order of a few percent.
 Due to the violation of temporal part of deSitter symmetry, 
 a deviation from the scale invariant spectrum can be expected to be of the order of the 
 slow roll parameters. Actually, this deviation has been observationally 
 confirmed~\cite{Komatsu:2010fb}.
 In this line of thought, it is legitimate to suspect that the spatial part of
 deSitter symmetry also breaks down slightly.
 The violation of the spatial part of deSitter symmetry would lead to
 an anisotropy in the cosmic expansion, namely, anisotropic inflation. 
 As a consequence, quantum fluctuations generated during the anisotropic inflation
  must have the statistical anisotropy.  Therefore, 
it is quite natural to expect the statistical anisotropy of the order of 
the slow roll parameter.
 
 Historically, there have been many attempts to construct anisotropic inflationary 
 models~\cite{Ford:1989me,Kaloper:1991rw,Barrow:2005qv,Barrow:2009gx,Campanelli:2009tk,Golovnev:2008cf,Kanno:2008gn,Ackerman:2007nb}. However, it has been shown that these models 
suffer from the instability~\cite{Himmetoglu:2008zp}, or a fine tuning problem, 
or a naturalness problem. 
Recently, a successful anisotropic inflationary model has been
 proposed~\cite{Watanabe:2009ct,Kanno:2009ei}. More precisely, 
it turned out that the presence of a non-trivial gauge kinetic function in supergravity theory
 can accommodate an anisotropic inflation. 
 It is well known that the supergravity is characterized by a superpotential,
 a Kahler potential, and a gauge kinetic function.
 These functions should be constrained 
 by comparing predictions of inflation with cosmological observations.
 For example, the tilt of the power spectrum has given interesting information
for the superpotential  and the Kahler potential. 
 The information provides a hint to the fundamental theory.
Amazingly, so far, the gauge kinetic function in supergravity has been neglected 
in making predictions of inflation. 
The reason is partially due to the cosmic no-hair theorem
which states that the anisotropy, curvature, and any matter will vanish
once the inflation commences~\cite{Wald:1983ky}. 
It has been proved that this is merely a prejudice~\cite{Watanabe:2009ct,Kanno:2009ei}.
The reason is simply that we do not have a cosmological constant
because of the violation of the translation symmetry for the inflaton. 
 In spite of the absence of the cosmic no-hair theorem, 
 since the anisotropic inflation is an attractor, 
 the predictability of the model still 
 remains~\cite{Himmetoglu:2009mk,Dulaney:2010sq,Gumrukcuoglu:2010yc,Watanabe:2010fh}.
 Indeed, the imprints of the anisotropic expansion could be seen 
 in the CMB~\cite{Watanabe:2010bu}.
 
From the particle physics point of view,
it is important to explore the role of the gauge kinetic function
in inflation. In a previous work, we have considered an abelian 
gauge field~\cite{Watanabe:2009ct}.
However, in reality, we have non-abelian gauge fields in particle physics models.
Hence, the main purpose of this paper is to investigate a cosmological role of
non-abelian gauge fields in an inflationary scenario.
There are two important differences between abelian and non-abelian gauge fields,
that is, the non-abelian gauge fields have multi-gauge-components and nonlinear self-couplings.
Thus, in this paper, we focus on consequences stemming from these two features. 

Here, we should note other works on the statistical anisotropy. The statistical anisotropy
generated by vector fields is first investigated using $\delta N$ formalism
 in \cite{Yokoyama:2008xw} where the possibility that the anisotropy appears
 strongly only in the non-gaussianity is pointed out. The model has been  further extended 
 in various ways~\cite{Karciauskas:2008bc,Dimopoulos:2009am,Dimopoulos:2009vu,ValenzuelaToledo:2009af,
ValenzuelaToledo:2009nq}. In particular, the formalism has been
 generalized to non-abelian gauge models~\cite{Bartolo:2009pa,Bartolo:2009kg,
 Dimastrogiovanni:2010sm}. As a different approach, there are attempts
  to see the remnant of the universe
before inflation~\cite{Tseng:2009xw,BlancoPillado:2010uw,Adamek:2010sg,Gumrukcuoglu:2006xj,Pitrou:2008gk,Gumrukcuoglu:2008gi}.
This could be possible if the duration of inflation is sufficiently short.
In the non-inflationary scenario, there is another mechanism for producing the statistical
anisotropy~\cite{Libanov:2010nk}. 

The organization of the paper is as follows.
In section II, we introduce inflationary models inspired by supergravity
where the Yang-Mills field couples with an inflaton.
In section III, we first solve basic equations numerically
and obtain solutions which exhibit anisotropic expansion and chaos during reheating. 
Next, we present analytical formula for the degree of the anisotropy of the
cosmic expansion during inflation. 
We also discuss observational implication of our finding. 
In section IV, we calculate a Lyapunov exponent of the chaos during reheating
and find no correlation between the anisotropy and the Lyapunov exponent. 
 The final section is devoted to conclusion. In the appendix A, we explain
how to reduce the degree of freedom of
the non-abelian gauge fields using the symmetry in the system. 

\section{Inflation model in supergravity}

In this section, we present an inflationary model based on supergravity
where we have a non-trivial gauge kinetic function for a gauge field. 
Although the gauge group could be general,
we choose $SU(2)$ for concreteness. Using Pauli matrices $\sigma^a$,
we can define generators of $SU(2)$ by $T^a=\sigma^a/2$ ($a=1,2,3$) satisfying
\begin{equation}
 [T^a,T^b]=i\epsilon^{abc}T^c\ ,\quad
 \textrm{tr}(T^a T^b)=\frac{1}{2}\delta^{ab} \ ,
\end{equation} 
where $\epsilon^{abc}$ is a Levi-Civita symbol and $\textrm{tr}$ denotes the trace
of the matrix representation. Here, $\delta^{ab}$ is a usual Kronecker delta.
The $SU(2)$ gauge field is defined as $A=A_\mu dx^\mu =A^a_\mu T^a dx^\mu$.
We note that the gauge field $A= A^a T^a$ has multi-gauge-component $A^a$.

The action for the gravitational field, the inflaton $\phi$, and the gauge field reads
\begin{equation}
 S=\int d^4x\sqrt{-g}\left[
\frac{1}{2\kappa^2}R-\frac{1}{2}(\partial\phi)^2
-V(\phi)-\frac{1}{2}f^2(\phi)\textrm{tr}(F_{\mu\nu} F^{\mu\nu})
\right] \ ,
\label{action}
\end{equation}
where $R$ is the scalar curvature, $g$ represents a determinant of the spacetime metric,
 $V(\phi)$ is a potential for the inflaton and the field
strength $F_{\mu\nu}$ of the $SU(2)$-gauge field is defined as
$F_{\mu\nu}=\partial_\mu A_{\nu}-\partial_\nu A_{\mu}+ig_{Y} [A_\mu,A_\nu]$.
Here, $g_Y$ is a Yang-Mills coupling constant.
The above action is invariant under the local $SU(2)$ gauge transformation,
\begin{equation}
 A_{\mu}\to \gamma^{-1}A_\mu \gamma -\frac{i}{g_{Y}}\gamma^{-1}\partial_\mu
  \gamma\ ,
\end{equation}
where $\gamma\in SU(2)$. The gauge kinetic function $f(\phi)$ will be specified later.
Equations of motion derived from the action~(\ref{action}) are given by
\begin{eqnarray}
&&\frac{1}{\kappa^2}G_{\mu\nu}
=\nabla_\mu\phi
\nabla_\nu\phi-\frac{1}{2}g_{\mu\nu}(\nabla\phi)^2-g_{\mu\nu}V(\phi)
+2f^2(\phi)\,\textrm{tr}[F_{\mu\rho}F_\nu{}^\rho-\frac{1}{4}g_{\mu\nu}F^2]\
,\label{Eeq}\\
&&\nabla^2\phi-V'(\phi)-f(\phi)f'(\phi)\,\textrm{tr}(F^2)=0\ ,\\
&&D_\nu[f^2(\phi)F^{\mu\nu}]=0\ ,\label{gauge_eq}
\end{eqnarray}
where $G_{\mu\nu}$ is the Einstein tensor, $\nabla_\mu$ represents
a covariant derivative with respect to the metric $g_{\mu\nu}$ and
we have defined the derivative $'\equiv d/d\phi$ and the gauge covariant derivative
$D_\mu=\nabla_\mu+ig_{Y} [A_\mu,\ \ast\ ]$.

Now, let us consider a cosmological background spacetime. 
For simplicity, we consider the axially symmetric Bianchi type-I metric 
\begin{equation}
 ds^2=-dt^2+e^{2\alpha(t)}[e^{-4\sigma(t)}dx^2+e^{2\sigma(t)}(dy^2+dz^2)]
 \ ,
\label{Imetric}
\end{equation}
where $\alpha$ describes the average expansion and $\sigma$ characterizes the anisotropy
of the expansion. 
The symmetry  in this spacetime is characterized by Killing vectors
$\partial_x$, $\partial_y$, $\partial_z$ and
$\xi_\phi\equiv-z\partial_y+y\partial_z$. In particular, $\xi_\phi$ generates the
rotational symmetry in $(y,z)$-plane. 
Imposing the symmetry on the inflaton $\phi$ and the gauge field $A$, we can reduce
variables into the following form
\begin{equation}
 \phi(x^\mu)=\phi(t)\ ,\quad
 A(x^\mu)=v_1(t) T^1 dx + v_2(t) (T^2 dy + T^3 dz)\ . 
\label{sym_variable}
\end{equation}
The gauge field $A$ is parametrized by two functions, $v_1(t)$ and $v_2(t)$.
In the appendix \ref{app:Axisym}, we explain 
how to achieve the above form for the gauge field by using the symmetry.
Substituting Eqs.(\ref{Imetric}) and (\ref{sym_variable}) into
Eqs.(\ref{Eeq}-\ref{gauge_eq}), 
we obtain basic equations for the cosmological background spacetime. 
From the time-time component of Einstein equations, we obtain a constraint equation
\begin{eqnarray}
&&\frac{3}{\kappa^2}(-\dot{\alpha}^2+\dot{\sigma}^2)
+\frac{1}{2}\dot{\phi}^2+V(\phi) \nonumber\\
&&\hspace{1cm}+\frac{1}{2}f^2(
e^{-2\alpha+4\sigma}\dot{v}_1^2
+2e^{-2\alpha-2\sigma}\dot{v}_2^2
+2 g_{Y}^2 e^{-4\alpha+2\sigma}v_1^2v_2^2
+ g_{Y}^2 e^{-4\alpha-4\sigma}v_2^4)=0 \ ,
\label{constraint}
\end{eqnarray}
where we defined a derivative $\cdot\equiv d/dt$ with respect to the cosmic time. 
We also have the evolution equations
\begin{eqnarray}
&&\frac{2}{\kappa^2}\ddot{\alpha}+\frac{3}{\kappa^2}(\dot{\alpha}^2+\dot{\sigma}^2)
+\frac{1}{2}\dot{\phi}^2-V \nonumber\\
&&\hspace{1cm}
+\frac{1}{6}f^2(
e^{-2\alpha+4\sigma}\dot{v_1}^2+2e^{-2\alpha-2\sigma}\dot{v_2}^2
+2 g_{Y}^2 e^{-4\alpha+2\sigma}v_1^2v_2^2
+ g_{Y}^2 e^{-4\alpha-4\sigma}v_2^4
)=0  
\label{alphaeq}
\end{eqnarray}
and
\begin{eqnarray}
\ddot{\sigma}+3\dot{\alpha}\dot{\sigma}
-\frac{\kappa^2}{3}f^2(
e^{-2\alpha+4\sigma}\dot{v_1}^2-e^{-2\alpha-2\sigma}\dot{v_2}^2
- g_{Y}^2 e^{-4\alpha+2\sigma}v_1^2v_2^2
+ g_{Y}^2 e^{-4\alpha-4\sigma}v_2^4
)=0 \ .
\label{sigmaeq}
\end{eqnarray}
The equation for the inflaton yields
\begin{eqnarray}
\ddot{\phi}+3\dot{\alpha}\dot{\phi}+V'
-ff'(e^{-2\alpha+4\sigma}\dot{v_1}^2+2e^{-2\alpha-2\sigma}\dot{v_2}^2-2
g_{Y}^2 e^{-4\alpha+2\sigma}v_1^2v_2^2- g_{Y}^2 e^{-4\alpha-4\sigma}v_2^4)=0
\ .
\label{phieq}
\end{eqnarray}
From Yang-Mills equations, we have
\begin{eqnarray}
\ddot{v_1}+2\frac{f'}{f}\dot{\phi}\dot{v_1}+(\dot{\alpha}+4\dot{\sigma})\dot{v_1}
+2 g_{Y}^2 e^{-2\alpha-2\sigma}v_1v_2^2=0
\label{v1eq}
\end{eqnarray}
and
\begin{eqnarray}
\ddot{v_2}+2\frac{f'}{f}\dot{\phi}\dot{v_2}+(\dot{\alpha}-2\dot{\sigma})\dot{v_2}
+ g_{Y}^2 e^{-2\alpha+4\sigma}v_1^2v_2+ g_{Y}^2 e^{-2\alpha-2\sigma}v_2^3=0 \ .
\label{v2eq}
\end{eqnarray}
Note that, when we put $v_2 =0$, the above equations reduce to those in abelian cases.

We need to specify the
 inflaton potential $V(\phi)$ and the coupling function $f(\phi)$
in order to solve the above equations. 
From the previous analysis in \cite{Watanabe:2009ct}, we know that the condition
\begin{eqnarray}
  \frac{f'}{f} \frac{V'}{V} > 2 \kappa^2 \ .
  \label{condition}
\end{eqnarray}
is necessary for anisotropic inflation to commence.
When we consider a simple chaotic inflation
\begin{equation}
 V(\phi)=\frac{1}{2}m^2\phi^2\ ,
\end{equation}
the simplest choice is~\cite{Martin:2007ue}
\begin{equation}
 f(\phi)=e^{c\kappa^2\phi^2/2} \ .
\end{equation}
Then, the condition (\ref{condition}) for anisotropic inflation yields $c>1$.
It should be stressed that the anisotropic inflation occurs
for a quite broad class of potential and gauge kinetic functions
as long as the condition (\ref{condition}) is 
satisfied~\cite{Moniz:2010cm,Kanno:2010nr,Emami:2010rm,Dimopoulos:2010xq}. 

In the next section, we solve Eqs.(\ref{constraint}-\ref{v2eq}) and
study the anisotropic inflation caused by the gauge kinetic
function for the $SU(2)$ Yang-Mills field.

\section{Anisotropic inflation}

Because of the non-linearity of basic equations of motion, it is difficult
to obtain exact solutions. Hence, we first solve equations of motion numerically
and find features of anisotropic inflation. From the numerical analysis,
it tuns out that the nonlinearity of gauge fields
can be neglected during anisotropic inflation.
Thus, we can make an analytic treatment of basic equations of motion
under the slow roll approximation. 

\subsection{Numerical analysis}

Let us solve the Eqs.(\ref{alphaeq}-\ref{v2eq}) numerically.
In our numerical calculations, we set parameters as 
\begin{equation}
\kappa=1,\quad c=2,\quad g_{Y}=0.01,\quad m=10^{-5}  \ ,
\end{equation}
and initial conditions as
\begin{equation}
 \phi=12,\quad
 \dot{\phi}=0,\quad 
v_1=0,\quad 
 \dot{v}_1=2.47\times 10^{-75},\quad
v_2=0,\quad
 \alpha=\sigma=\dot{\sigma}=0\ .
\label{inic}
\end{equation}
The initial value for $\dot{\alpha}$ is determined
 by the constraint equation~(\ref{constraint}).
We use these parameters for all numerical calculations in this paper
since the qualitative result does not change even if we change these parameters. 
However, the initial condition for $\dot{v}_2$, which is not included
 in the above set (\ref{inic}), 
changes the qualitative behaviour of the inflation.
Thus, in our numerical calculations, 
we vary the value of $\dot{v}_2$ 
and study features of anisotropic inflation for various $\dot{v}_2$'s.

\begin{figure}[h]
\begin{center}
\includegraphics[height=9cm,clip,angle=270]{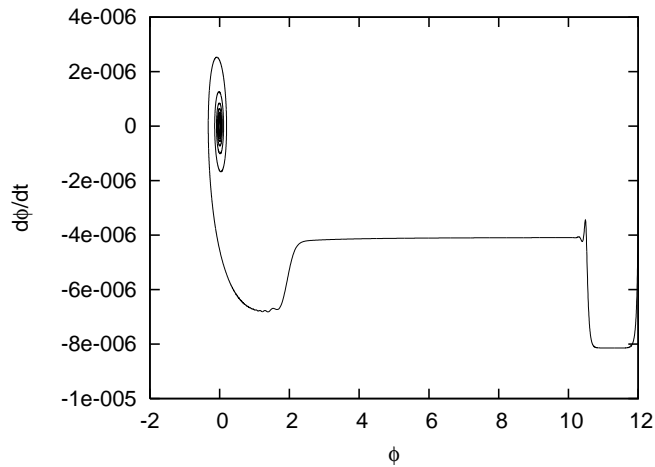}
\caption{\label{fig:phase} 
Te phase flow in $\dot{\phi}$-$\phi$
space for $\dot{v}_2/\dot{v}_1=0.5$.
We can see two phases of inflations which correspond to 
isotropic and anisotropic inflations.
}
\end{center}
\end{figure}
In Fig.\ref{fig:phase}, we show a trajectory with the initial condition
$\dot{v}_2/\dot{v}_1=0.5$ in $\dot{\phi}$-$\phi$ plane.
Taking look at Fig.\ref{fig:phase}, we see that the behaviour is similar to
that in an anisotropic inflation in the case of the 
$U(1)$-gauge field~\cite{Watanabe:2009ct}. 
There are two phases of inflations, isotropic and
anisotropic inflations. Since we have started with negligible vector fields,
the trajectory goes into a conventional isotropic inflation. 
During this stage, the energy density of the vector fields rapidly increases
and the inflation soon becomes anisotropic in the second slow roll stage.
There, the increase of the energy density of the vector fields saturates
due to the backreaction of the vector field. 
Therefore, the anisotropic inflation is an attractor solution.
After a sufficient e-folding, the inflation ends with reheating. 

\begin{figure}[h]
\begin{center}
\includegraphics[height=9cm,clip,angle=270]{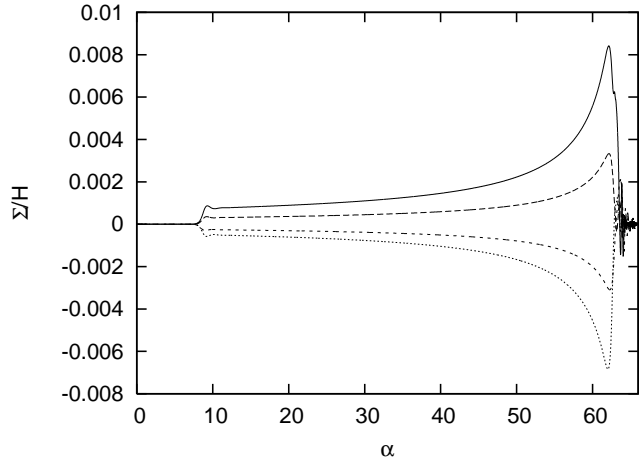}
\caption{\label{fig:aniso}
Time evolution of anisotropy $\Sigma/H$ is plotted against e-holding
 number $\alpha$. 
These curves correspond to $\dot{v}_2/\dot{v}_1=0.5$, $0.75$, $1.33$ and $2.0$ from top to bottom. 
We see that the anisotropy can be either positive or negative
depending on the ratio $\dot{v}_2/\dot{v}_1$. We also see the rapid oscillation
of the anisotropy during the reheating.
}
\end{center}
\end{figure}
\begin{figure}[h]
\begin{center}
\includegraphics[height=9cm,clip,angle=270]{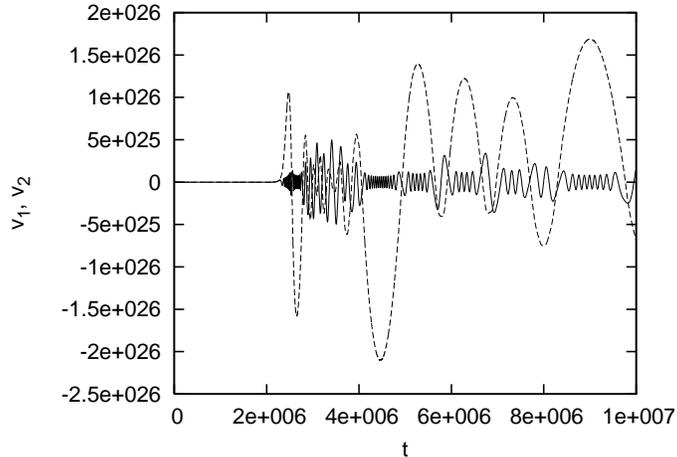}
\caption{\label{fig:chaos} 
The dynamics of $v_1(t)$ and $v_2(t)$ with the initial condition
 $\dot{v}_2/\dot{v}_1=0.5$.
The dashed and solid curves correspond to $v_1(t)$ and $v_2(t)$.
The horizontal axis is the cosmological time $t$ in the unit of $\kappa=1$.
We can see the chaotic behaviour of the gauge field after inflation.
}
\end{center}
\end{figure}

In Fig.\ref{fig:aniso}, we show time evolution of anisotropy,
$\Sigma/H=\dot{\sigma}/\dot{\alpha}$, for $\dot{v}_2/\dot{v}_1=0.5$, $0.75$, $1.33$ and $2.0$.
It is remarkable that the anisotropy can be either positive or negative
depending on the initial ratio $\dot{v}_2/\dot{v}_1$. On the other hand, 
for the anisotropic inflation by a $U(1)$-gauge field, 
the anisotropy did not depend on the initial
condition for the gauge field and it was always positive.
In~\cite{Watanabe:2010fh}, 
it was shown that the statistical anisotropy generated by anisotropic
inflation with a $U(1)$-gauge field has an opposite sign to the one
claimed by the analysis of WMAP data in the CMB.
The negative anisotropy, if true, may suggests
that the anisotropic inflation assisted by non-abelian gauge field generates the
statistical anisotropy. 
We should note that the massive vector fields could also produce the
negative anisotropy~\cite{Dimopoulos:2010xq}.
We note that the initial condition dependence
does not mean the loss of predictability because the initial condition 
dependence comes into only in the degree of the anisotropy and hence it
can be absorbed by rescaling the model parameter.
Indeed, the consistency relations among observables found in \cite{Watanabe:2010fh}
are independent on the initial conditions.  

In the reheating stage, we see the rapid oscillation of the anisotropy
in Fig.\ref{fig:aniso}, which implies a rapid oscillation of the gauge field. 
In Fig.\ref{fig:chaos}, we show the dynamics of the gauge field, 
$v_1(t)$ and $v_2(t)$ for $\dot{v}_2/\dot{v}_1=0.5$.
We can see the chaotic behaviour of the gauge field during the reheating.
We discuss the Lyapnov exponent which characterize the chaos 
in section.\ref{sec:chaos}.

\subsection{Slow roll approximation}
\label{sec:slow}

In the previous subsection, we found the slow roll inflation with the anisotropy
of the expansion of the universe. 
Now, we solve basic equations (\ref{constraint}-\ref{v2eq}) 
using the slow roll approximation.

In the inflationary phase, the scalar field takes the value  $\kappa\phi\sim 10$.
Then, the gauge kinetic function has a very large value, $f(\phi)\sim e^{100}$. 
As one can see from the action~(\ref{action}), 
the $g_{Y}/f(\phi)$ can be regarded as an effective gauge coupling.
During the inflation, the effective gauge coupling becomes very
small $g_{Y}/f(\phi)\sim e^{-100}$. 
Therefore, we can neglect the gauge coupling during inflation.
Then, we can integrate Eqs.(\ref{v1eq}) and (\ref{v2eq}) as
\begin{equation}
 \dot{v}_1=f^{-2}(\phi)e^{-\alpha-4\sigma}p_1\ ,\quad
 \dot{v}_2=f^{-2}(\phi)e^{-\alpha+2\sigma}p_2\ ,
\label{p1p2}
\end{equation}
where $p_1$ and $p_2$ are constants of integration.
Since the energy of the gauge field should be subdominant during inflation, 
we can ignore $\sigma$ in Eqs.(\ref{constraint}-\ref{alphaeq}).
Thus, using the slow roll conditions, $\dot{\phi}^2<<V$ and
$\ddot{\phi}<<V'$, these equation can be written as
\begin{eqnarray}
&&\dot{\alpha}^2=\frac{\kappa^2}{6}
[m^2\phi^2+e^{-c\kappa^2\phi^2-4\alpha}(p_1^2+2p_2^2)]
\label{slow_con0}
\\
&&3\dot{\alpha}\dot{\phi}+m^2\phi
-c\kappa^2\phi e^{-c\kappa^2\phi^2-4\alpha}(p_1^2+2p_2^2
)=0\ ,\label{slow_phi}
\end{eqnarray}
When the effect of the vector field is comparable with that of the
inflaton in (\ref{slow_phi}), namely, when
$m^2\sim c\kappa^2 e^{-c\kappa^2\phi^2-4\alpha}(p_1^2+2p_2^2)$, 
we find 
$e^{-c\kappa^2\phi^2-4\alpha}(p_1^2+2p_2^2)/(m^2\phi^2)\sim 1/(c\kappa^2\phi^2)\sim 10^{-2}$.
Thus, we can neglect the second term in the right hand side of
Eq.(\ref{slow_con0}). Then, from Eqs.(\ref{slow_con0}) and (\ref{slow_phi}), we find
\begin{equation}
\phi \frac{d\phi}{d\alpha}=
-\frac{2}{\kappa^2}
+\frac{2c}{m^2} e^{-c\kappa^2\phi^2-4\alpha}(p_1^2+2p_2^2
)\ ,
\end{equation}
Integrating the above equation, we obtain
$e^{-c\kappa^2\phi^2-4\alpha}=m^2(c-1)/(c^2\kappa^2(p_1^2+2p_2^2)(1+De^{-4(c-1)\alpha}))$,
where $D $ is a constant of integration. This solution rapidly converges to
\begin{equation}
 e^{-c\kappa^2\phi^2-4\alpha}=\frac{m^2(c-1)}{c^2\kappa^2(p_1^2+2p_2^2)}\ .
\label{slow1}
\end{equation}
From Eq.(\ref{sigmaeq}), we obtain 
$3\dot{\alpha}\dot{\sigma}=(\kappa^2/3)e^{-c\kappa^2\phi^2-4\alpha}(p_1^2-p_2^2)$.
Therefore, the anisotropy can be evaluated as
\begin{equation}
  \frac{\Sigma}{H}
=\frac{\kappa^2}{9\dot{\alpha}^2}e^{-c\kappa^2\phi^2-4\alpha}(
p_1^2-p_2^2
)
=\frac{2(c-1)(p_1^2-p_2^2)}{3c^2(p_1^2+2p_2^2)}\frac{1}{\kappa^2\phi^2}\ .
\end{equation}
To obtain the last expression, we have used Eqs.(\ref{slow_con0}) and
(\ref{slow1}).
From (\ref{constraint}) and (\ref{alphaeq}), we have 
$\ddot{\alpha}=-(\kappa^2/2)\dot{\phi}^2 -(\kappa^2/3)e^{-c\kappa^2\phi^2-4\alpha}(p_1^2+2p_2^2)$.
Thus, the slow roll parameter is given by
\begin{equation}
 \epsilon=-\frac{\ddot{\alpha}}{\dot{\alpha}^2}=\frac{2}{c\kappa^2\phi^2}
\end{equation}
Therefore, the anisotropy can be written as
\begin{equation}
  \frac{\Sigma}{H}
=\frac{(c-1)(p_1^2-p_2^2)}{3c(p_1^2+2p_2^2)}\epsilon\ .
\label{aniso_ep}
\end{equation}
We should notice that
 the abelian result can be recovered if we put $p_2 =0$~\cite{Watanabe:2009ct}.
From the expression (\ref{aniso_ep}), we obtain an inequality as
\begin{equation}
  -\frac{c-1}{6c}\epsilon \leq\frac{\Sigma}{H}\leq \frac{c-1}{3c}\epsilon\ .
\end{equation}
It implies that $\Sigma/H$ is suppressed by the slow roll parameter.
The anisotropy can be either positive or negative
depending on the ratio $p_2/p_1$. In particular, when $p_1= p_2$, we have
no anisotropy. Except for this accidental case, we have the anisotropy
proportional to the slow roll parameter. In particular, for the cases $p_2 > p_1$,
we have a negative anisotropy $\Sigma/H$ which never occurs in abelian models. 
This analytical result explains the numerical result shown in Fig.\ref{fig:aniso}.

The statistical anisotropy induced by an anisotropic inflation can be characterized
by the direction dependent power spectrum
\begin{eqnarray}
P ({\bf k}) = P_0 (k) \left[ 1 + g_* \left( {\bf k}\cdot {\bf n}\right)^2 \right] \ ,
\end{eqnarray}
where $P_0 (k)$ is the isotropic part of the power spectrum, 
${\bf k}$ is a wavenumber vector of fluctuations and 
${\bf n}$ is a specific direction, in our case, this is the $x$-direction.
Here, the number $g_*$ is a magnitude of the statistical anisotropy. 
According to the recent analysis of the CMB data~\cite{Groeneboom:2008fz},
   $g_*$ takes a positive value. 
 The prediction of anisotropic inflation based on the abelian gauge field models  
 was inconsistent with
 this result~\cite{Himmetoglu:2009mk,Dulaney:2010sq,Gumrukcuoglu:2010yc,Watanabe:2010fh}. 
In our non-abelian models, however, the anisotropy can give rise to a positive $g_*$
since the anisotropy of the expansion can take any signature depending
on the ratio $p_2/p_1$. Thus, if the CMB data shows the statistical anisotropy
with positive $g_*$, it may imply the anisotropic inflation with
a gauge kinetic function for a non-abelian gauge field.
Of course, we should keep it in mind that the data analysis may contain systematic
errors. 

\section{Chaos during reheating}
\label{sec:chaos}

In this section, we study the chaos during reheating in
Fig.\ref{fig:chaos}. The chaos in Yang-Mills cosmology has been studied previously
~\cite{Barrow:1997sb,Jin:2004vh,Darian:1996jf}. In those analysis, the existence of coherent
non-abelian gauge fields is assumed. Here, the initial coherent non-abelian
gauge field is provided by the anisotropic inflation.

As we explained at the beginning of subsection.\ref{sec:slow},
the gauge coupling is extremely weak during the inflation.
Hence, we can neglect the non-linearity of the gauge field.
During this phase, the coherent gauge field is produced due to the rapid variation
of the gauge kinetic function.
While, at the end of the inflation, the gauge kinetic function becomes
$f(\phi)\sim 1$ and effective gauge coupling $g_{Y}/f(\phi)$ becomes of the order of $g_{Y}$.
The chaotic behaviour after the inflation occurs
due to this relatively large effective gauge coupling.
Therefore, without the time varying gauge kinetic function, 
this chaotic behavior of the gauge field never happens.

\begin{figure}
\begin{center}
\includegraphics[height=9cm,clip,angle=270]{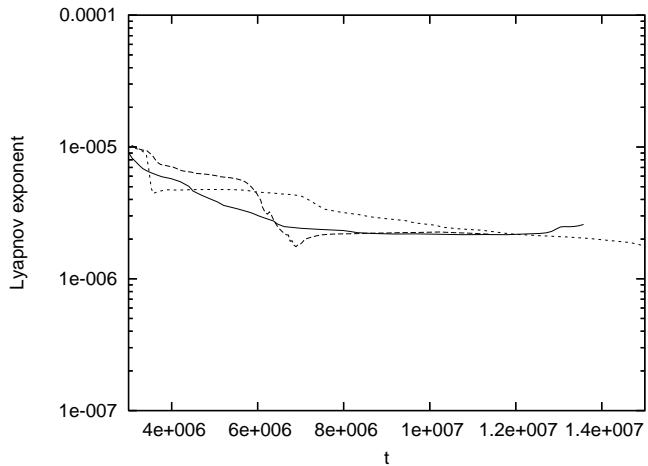}
\caption{\label{fig:Liap} 
The Lyapunov exponents $\lambda(t)$ for various initial conditions.
Solid, dashed and dotted curves correspond to
$\Sigma/(\epsilon H)=6.20\times 10^{-2}$,
$2.26\times 10^{-3}$ and
$-7.10\times 10^{-2}$, respectively.
We cannot find any correlation between the Lyapunov exponents.
}
\end{center}
\end{figure}

From the numerical analysis, 
the pattern of oscillation in Fig.\ref{fig:chaos} 
seems to change depending on the ratio $\dot{v}_2/\dot{v}_1$.
The anisotropy $\Sigma/(\epsilon H)$ also depends on the same ratio. 
Hence, we might have some relation between
the anisotropy and the chaos. 
So, we need to check if the relation between anisotropy and chaotic behaviour exists.
We consider linear perturbations of Eqs.(\ref{constraint}-\ref{v2eq}) by the substitution
$\alpha\to \alpha+\delta\alpha$, $\sigma\to\sigma+\delta\sigma$, 
$\phi\to\phi+\delta\phi$, $v_1\to v_1+\delta v_1$ and $v_2\to v_2+\delta v_2$.
Since the chaotic system is sensitive to initial conditions,
the perturbation would grow at late time.
We define ``Lyapunov exponent'' $\lambda(t)$ as
$\delta v_2 \propto e^{\lambda(t) t}$.\footnote{
Mathematically, the Lyapunov exponent $\lambda$ is defined by
$\lambda=\lim_{t\to \infty} \ln(|\bm{w}(t)|/|\bm{w}(0)|)/t$
where
$\bm{w}(t)=(\delta\alpha,\cdots,\delta v_2, \dot{\delta\alpha},\cdots,\dot{\delta v_2})$.
Since it is difficult to solve
Eqs.(\ref{constraint}-\ref{v2eq}) for sufficiently long time in our numerical calculations, 
we define the $\lambda(t)$ in this way.}
The $\lambda(t)$ represents the growing rate of the perturbation.
In Fig.\ref{fig:Liap}, we depict the $\lambda(t)$ for
$\Sigma/(\epsilon H)=6.20\times 10^{-2}$,
$2.26\times 10^{-3}$ and
$-7.10\times 10^{-2}$, respectively. We cannot find any
correlation between $\lambda(t)$ and $\Sigma/(\epsilon H)$.
It indicates that Lyapunov exponent does not depend on the anisotropy
in the anisotropic inflation. 
Although there is a gauge ambiguity in defining the Lyapunov exponent,
 our conclusion itself does not depend on the choice of the gauge.

\section{Conclusion}

We have studied the anisotropic inflation model inspired by supergravity where the 
gauge kinetic function for the $SU(2)$ Yang-Mills field is non-trivial.
We found that the anisotropy of expansion rate can take either positive or negative value
depending on the initial ratio $\dot{v}_2/\dot{v}_1$. 
Namely, the shape of the comoving volume becomes either prolate or oblate
depending on the initial configurations of the gauge field.
This new feature can be attributed to the multi-component nature of non-abelian gauge fields.
In principle, this could occur even in multi-abelian models. However,
it would be difficult to organize the multi-abelian fields so that the
same result as the non-abelian models can be obtained. On the other hand,
the gauge structure of the non-abelian gauge field self-organizes the configuration.
In spite of the above initial configuration dependence, 
the anisotropic inflation is still an attractor in the sense that
the details of initial  conditions are irrelevant except for one relevant parameter
which controls the anisotropy of the universe. 
In addition to the anisotropic expansion, 
we found the chaotic behaviour of the gauge field during reheating.
This is due to the nonlinear self-coupling of non-abelian gauge fields.
We calculated the Lyapunov exponent of the chaos and found that the Lyapunov
exponent does not correlate with the anisotropy of the inflation.
This indicates the universality of the chaotic behaviour of Yang-Mills field
in an anisotropic inflationary scenario with a gauge kinetic function
for a non-abelian gauge field.

It is remarkable that the anisotropy $\Sigma/H$ can be negative in our inflation
model. In~\cite{Watanabe:2010fh}, 
it was shown that statistical anisotropy generated by anisotropic
inflation by $U(1)$-gauge field takes an opposite sign to the one of
observationally favored value. While, the negative anisotropy would lead to
the observationally favored signature. It may indicate that the anisotropic 
inflation with a gauge kinetic function for a non-abelian gauge field 
generates the statistical anisotropy in the present CMB.
Probably, it would be too early to conclude something, at least,
we should wait for more precise CMB data provided by PLANCK~\cite{Ma:2011ii}.
It is also intriguing to observe that the new relevant parameter $\dot{v}_2/\dot{v}_1$
helps to
weaken the constraint on the model parameter $c$ found in \cite{Watanabe:2010fh}.
This is because we have two parameters $\dot{v}_2/\dot{v}_1$ and $c$
to control the anisotropy. 

In this paper, we have assumed axial symmetry for simplicity.
It is a straightforward exercise to consider more general configurations.
As a future work, we can consider non-gaussianity in an anisotropic inflation
along the previous paper~\cite{Yokoyama:2008xw}. 
From this point of view, it is interesting to extend anisotropic inflationary
scenario to a non-slow roll type inflationary scenario
 such as the DBI inflation model~\cite{Silverstein:2003hf}. 
It is also interesting to study implications of chaos in the early universe.
For example, gravitational waves may be generated at a reheating
phase. Because of the chaotic behaviour of the gauge field, we may be able
to find relic of the chaos during the reheating.

\noindent
{\it Acknowledgements}\\
JS would like to thank Sugumi Kanno and Masa-aki Watanabe for previous collaboration
and useful comments.
We are grateful to Kazuya Koyama and David Wands for the hospitality
during our stay at ICG, the University of Portsmouth , 
where most of this work has been done. 
KM is supported by a grant for research abroad by the JSPS (Japan).
JS is supported by the Grant-in-Aid
for Scientific Research (A) (No.21244033),  the
Grant-in-Aid for  Scientific Research Fund of the Ministry of
Education, Science and Culture of Japan No.22540274, the Grant-in-Aid
for Scientific Research (A) ( No. 22244030), the
Grant-in-Aid for  Scientific Research on Innovative Area No.21111006,
JSPS under the Japan-Russia Research Cooperative
Program and the Grant-in-Aid for the Global COE Program
``The Next Generation of Physics, Spun from Universality and Emergence''.

\appendix

\section{Axi-symmetric gauge fields}
\label{app:Axisym}

\subsection{Symmetry constraints}

Using symmetry, we can make a reduction of variables~\cite{Darian:1996jf}.
In the present system, there are spacetime isometry and gauge symmetry.
Using these symmetry, we will make the variables as simple as possible. 

Imposing the translation invariance along $\partial_x$, $\partial_y$ and
$\partial_z$, we can put the gauge field as
$A=A_t(t)dt+A_i(t)dx^i$. 
Furthermore, using the local $SU(2)$ gauge freedom,  we can fix the time
component of the gauge field as $A_t(t)=0$. 
Then, the gauge field can be written as
\begin{equation}
 A^a=\beta^a(t)dx+\gamma^a(t)dy+\delta^a(t)dz\ .
\end{equation}
The residual global gauge transformation is given by
\begin{equation}
\begin{split}
 \delta_g A^a &= i[A,u]^a=\epsilon^{abc}u^bA^c\\
&=\epsilon^{abc}u^b(\beta^c(t)dx+\gamma^c(t)dy+\delta^c(t)dz)\\
&=(\vec{u}\times\vec{\beta})^a dx+(\vec{u}\times\vec{\gamma})^a
 dy+(\vec{u}\times\vec{\delta})^a dz\ .
\end{split}
\label{gauge}
\end{equation}
where $u^a$ are constants and we used a vector notation like as 
$\vec{u}=(u^1,u^2,u^3)$, $\vec{\beta}=(\beta^1,\beta^2,\beta^3)$, etc.
Without loss of generality, choosing an appropriate basis in Lie algebra, we can put
$\vec{u}=(u^1,0,0)=u^1\vec{e}_1$.

Now, we consider the rotational symmetry.
The rotational transformation is generated by 
$\mathcal{L}_\phi dx=0$, 
$\mathcal{L}_\phi dy=-dz$ and 
$\mathcal{L}_\phi dz=dy$. 
Under the infinitesimal rotational transformation,  the gauge field
transforms as
\begin{equation}
 \mathcal{L}_\phi A^a=-\gamma^a(t)dz+\delta^a(t)dy\ .
\label{rotA}
\end{equation}
For the rotational invariance, the above rotational transformation must be
absorbed by the gauge transformation~(\ref{gauge}), namely,
\begin{equation}
 \mathcal{L}_\phi A^a=\delta_g A^a. 
\label{local}
\end{equation}
Since the gauge field has to be the same after making a round,
$A= \exp(2\pi\mathcal{L}_\phi)A$ holds. 
Thus, using Eq.(\ref{local}) and $[\mathcal{L}_\phi,\delta_g]=0$,
we derive a relation
\begin{equation}
A= \exp(2\pi\mathcal{L}_\phi)A=\exp(2\pi \delta_g)A\ .
\end{equation}
Therefore, $\exp(2\pi \delta_g)$ must be an identical transformation.
This implies $u^1=n\in \bm{Z}$.
Then, substituting Eqs.(\ref{gauge}) and (\ref{rotA}) into Eq.(\ref{local}), we obtain
\begin{eqnarray}
 &&n\vec{e_1}\times\vec{\beta}=0\ ,\label{veceq1}\\
 &&n\vec{e_1}\times\vec{\gamma}=\vec{\delta}\ ,\label{veceq2}\\
 &&n\vec{e_1}\times\vec{\delta}=-\vec{\gamma}\ .\label{veceq3}
\end{eqnarray}
For $n=0$, we find
\begin{equation}
 \vec{\beta}=(\beta^1(t),\beta^2(t),\beta^3(t))\ ,\quad
 \vec{\gamma}=\vec{\delta}=0\ .
\label{n=0_case}
\end{equation}
For $n=1$, from Eq.(\ref{veceq1}), we obtain $\vec{\beta}\parallel \vec{e}_1$.
From Eqs.(\ref{veceq2}) and (\ref{veceq3}), 
we find $\vec{e}_1\perp\vec{\gamma}\perp\vec{\delta}\perp\vec{e}_1$.
Furthermore, taking the absolute value in Eq.(\ref{veceq2}), we have $|\vec{\gamma}|=|\vec{\delta}|$.
Thus, the vectors $\vec{\beta}$, $\vec{\gamma}$ and $\vec{\delta}$ can be written as
\begin{equation}
 \vec{\beta}=(\beta^1(t),0,0)\ ,\quad 
 \vec{\gamma}=(0,\gamma^2(t),\gamma^3(t))\ ,\quad 
 \vec{\delta}=(0,-\gamma^3(t),\gamma^2(t))\ .
\label{n=1_case}
\end{equation}
For $n>1$, Eqs.(\ref{veceq2}) and (\ref{veceq3}) cannot be
satisfied unless $\vec{\gamma}= \vec{\delta}=0$. Thus, 
we obtain
\begin{equation}
  \vec{\beta}=(\beta^1(t),0,0)\ ,\quad 
 \vec{\gamma}= \vec{\delta}=0\ .
\label{n>0_case}
\end{equation}


\subsection{Yang-Mills constraints}

In the previous subsection, we classified the gauge field 
into the three types (\ref{n=0_case}), (\ref{n=1_case}) and (\ref{n>0_case}). 
Here, we impose the Yang-Mills constraints on these expressions.

First, we consider the case (\ref{n=0_case}).
Substituting Eq.(\ref{n=0_case}) into a time component of Eq.(\ref{gauge_eq}),
we obtain
\begin{equation}
 \dot{\beta}^1\beta^2-\dot{\beta}^2\beta^1=
 \dot{\beta}^2\beta^3-\dot{\beta}^3\beta^2=
 \dot{\beta}^3\beta^1-\dot{\beta}^1\beta^3=0
\ .
\end{equation}
Therefore, we have $\vec{\beta}=\beta_1(t)\vec{c}$, where $\vec{c}$ is a constant
vector. Thus, choosing an appropriate basis in Lie algebra,
we can put
\begin{equation}
  \vec{\beta}=(\beta^1(t),0,0)\ ,\quad
 \vec{\gamma}=\vec{\delta}=0\ .
\label{n=0_2}
\end{equation} 
Next, we consider the case (\ref{n=1_case}). 
Then, from a time component of Eq.(\ref{gauge_eq}),
we find
\begin{equation}
  \dot{\gamma}^2\gamma^3-\dot{\gamma}^3\gamma^2=0\ .
\end{equation}
Thus, we get $\gamma^2=c\gamma^3$, where $c$ is a constant.
Choosing an appropriate basis in the subspace of Lie algebra spanned by 
$T^2$ and $T^3$, the vectors $\vec{\beta}$,
$\vec{\gamma}$ and $\vec{\delta}$ can be written as
\begin{equation}
 \vec{\beta}=(\beta^1(t),0,0)\ ,\quad 
 \vec{\gamma}=(0,\gamma^2(t),0)\ ,\quad 
 \vec{\delta}=(0,0,\gamma^2(t))\ .
\label{n=1_2}
\end{equation}
For the case (\ref{n>0_case}), constraint equations in Eq.(\ref{gauge_eq}) are
trivially satisfied and we have the same expression as Eq.(\ref{n=0_2}).
Therefore, Eq.(\ref{n=1_2}) is the most general expression for an axially
symmetric gauge field.

\end{document}